# Comment: Struggles with Survey Weighting and Regression Modeling

F. Jay Breidt and Jean D. Opsomer

We congratulate the author on an informative and thought-provoking discussion on a topic of broad interest to the statistics community: the fitting of models to data collected through complex surveys. The number of papers written on this topic, whether from a model-based or design-based perspective, is substantial and goes back at least to Konijn (1962). This topic has led to some disagreements between those advocating that the design best be ignored when the primary interest is on the characteristics of the model, and those stating that the design cannot be ignored. More recently, both sides of this discussion have moved to something approaching a consensus, with those favoring a model-based approach acknowledging the need to account for nonignorable designs in the model fitting, while the traditional design-based view has been extended to explore certain circumstances under which it is appropriate to ignore the design.

The current article is an excellent example of those recent discussions of why the design needs to be accounted for in modeling, and how this can be done in practice. The importance of *fully* accounting for the design by incorporating all relevant interactions provides a good motivation for the discussion of the range of methods in the article. It also stresses other aspects of importance to people working with survey data, in particular the desirability of maintaining scale/location invariance and linearity of the model-based estimators. This ensures consistency of estimates for different variables in the survey, as well as additivity over domains within the population. (As an aside, the poststratified estimator arising from logistic regression in Section 3.2 can be modified to yield approximate weights by the method proposed in Wu and Sitter, 2001.)

The article mentions a number of disadvantages of design-based (weighted) model fitting and inference. Weights are viewed as complicated and mysterious, in the sense that the modeler often does not know how they were constructed and hence might not want to rely on them when it comes to model specification and estimation. Estimation, and especially variance estimation, are viewed as more cumbersome under the design-based paradigm compared to a model-based analysis. In what follows, we will argue that a weighted analysis offers some distinct advantages and might actually reduce the complexity of the analysis in many cases, at least from the perspective of a statistician interested in using previously collected and weighted survey data to fit a model.

A key feature of the design-based paradigm (broadly speaking) is that it makes it possible to separate design and postsample adjustments from data analysis. Individuals tasked with creating survey weights are typically within the organization collecting the data, and will be referred here as "the survey statisticians." They have knowledge of the sampling design and have access to detailed information on the nonresponse characteristics of the sample and to relevant auxiliary information. Based on these sources of information, they develop a set of survey weights (and sometimes also produce sets of replication weights for variance estimation). As noted in the article, these weights are often much more complicated than simple inverses of inclusion probabilities, and in fact reflect the best effort on the part of the survey statisticians creating the weights to account for nonresponse and incorporate potentially useful population-level information. These weights are appended to the dataset, which is then made available to individuals interested in analyzing those data.

*F. Jay Breidt is Professor and Chair, Department of Statistics, Colorado State University, Fort Collins, Colorado 80523, USA e-mail: jbreidt@stat.colostate.edu. Jean D. Opsomer is Professor, Department of Statistics, Iowa State University, Ames, Iowa 50011, USA e-mail: jopsomer@iastate.edu.*

These individuals will be referred to as "the data analysts."

From the perspective of the data analysts, using these weights is convenient in the sense that they provide a simple way to account for the way the data were obtained, without requiring the data analysts to replicate many of the tasks of the survey statisticians. Overall, this "division of labor" allows both sets of statisticians to focus their efforts on the portion of the overall problem of most immediate interest to them, and for which they have both the expertise and the information available to best perform the required tasks.

As noted by a number of authors (e.g., Pfeffermann, 1993), performing a weighted analysis for a model using inverses of the inclusion probabilities ensures that the resulting estimators are design consistent for population-level quantities, which are themselves model consistent for the model parameters of interest. When the weights also include nonresponse adjustments (usually by way of poststratification) as well as other calibration adjustments, results for descriptive statistics, including those discussed in Särndal and Lundström (2005), show that the estimators are consistent under the joint design-response mechanism. While these results are expected to continue to hold when model parameters are targeted rather than finite population means, there is currently only limited formal theory exploring this topic.

The division of labor between the survey statisticians and the data analysts has some additional advantages. While the former typically have access to detailed unit-level information and can use that information in the construction of the weights, confidentiality issues often preclude such access for the latter. For instance, in the Social Indicators Survey considered in the Gelman article, avoiding the weights required knowledge of the number of adults and the number of phone lines in the household of each respondent, as well as various other demographic variables. It is easy to envision situations where at least some of these variables are not made available to the data analysts in order to protect the confidentiality of the survey respondents. In such situations, the data analysts could still try to build a model that incorporates the design effects, but might end up only being partly successful because some influential variables are not available.

Another consideration is the fact that large-scale surveys often involve complex stratification and post-stratification schemes, multiple phases and/or stages of selection, imputation for item nonresponse, etc. Accounting for all these factors, even if the needed sources of information are available to the data analysts, would require significant time and effort on the part of the data analysts and result in models that might be unwieldy and difficult to interpret.

One point noted in the Gelman article is that variance estimation for weighted estimators is more cumbersome than for fully model-based estimators. To a large extent, this is indeed the case, but a number of solutions are available. For specific models (e.g., linear or logistic regression), commercial software programs such as SAS are increasingly providing design-based estimation procedures, so that with access to the weights and some basic information about the design (e.g., stratification information and primary sampling unit identifiers), it is possible for the data analysts to perform design-based inference for model parameter estimators. An alternative procedure, already alluded to earlier and often used for large-scale surveys, is for the survey statisticians to provide sets of replication weights (e.g., jackknife or bootstrap replicates). In that case, variance estimation for the weighted estimates is a simple matter of recomputing the estimates for each set of replicate weights and calculating the variability among the replicate estimates.

Incorporating the design and nonresponse characteristics of a dataset through explicit modeling is a statistically valid and conceptually attractive approach to solving the nonignorability problem. It has the advantage of being easily integrated into the set of tools most familiar to data analysts, but, as explained in this interesting article, it requires knowledge of the relevant variables and has to be done carefully. Performing a design-based analysis with the weights provided as part of a survey dataset is attractive as well, because it is generally applicable even without detailed knowledge of the way the data were obtained.

In closing, we would like to suggest a number of possible developments that would help make data analysts more comfortable with these weighted analyses. While weight construction is likely to remain to a large extent an "art," more transparency in how weights are constructed might alleviate some of the discomfort on the part of data analysts having to rely on the work of survey statisticians as a building block in their own analysis. A related development might be more education and training in the interpretation of results of weighted analyses for nonsurvey statisticians and in methods for doing inference for design-weighted model estimates. On the survey statistics side, we would like to encourage the investigation of the statistical properties of weighted estimators for model parameters that explicitly accounts for the multiple adjustments typically made to survey weights, including calibration and nonresponse weighting.